\documentclass[letter]{aa}
\usepackage{txfonts}
\usepackage{graphicx,subfigure}
\usepackage{epsfig}
\usepackage{epsf}
\usepackage{lscape}
\usepackage{natbib}
\usepackage{rotating}
\usepackage{tabularx}
\usepackage{lscape}
\usepackage{supertabular}
\newcommand{\mic}{\,{\rm \mu m} }

\begin{document}
\title{Variations of the spectral index of dust emissivity from Hi-GAL
  observations of the Galactic plane\thanks{{\it Herschel} is an ESA space observatory
  with science instruments provided by European-led Principal
  investigator consortia and with important participation from NASA.}}

\author{D. Paradis \inst{1} \and M. Veneziani \inst{1,2} \\
\and 
A. Noriega-Crespo \inst{1} \and R. Paladini \inst{1} \and
F. Piacentini \inst{2} \and
J.P. Bernard \inst{3,4} \and P. de Bernardis \inst{2} \and L. Calzoletti
\inst{5} \and F. Faustini \inst{5}
\and P. Martin \inst{6} \and S. Masi \inst{2} \and L. Montier \inst{3,4} \and P. Natoli \inst{7} 
\and I. Ristorcelli \inst{3,4} \and M. A. Thompson \inst{8} \and
A. Traficante \inst{7}  \and  S. Molinari \inst{9}} 
 
\institute{Science Center, California Institute of Technology, Pasadena, CA
    91125, USA 
\and
Dipartimento di Fisica, Universita di Roma 1 La Sapienza, 00185 Roma, Italy 
\and
Universit\'e de Toulouse, UPS, CESR, 9 avenue du Colonel Roche,
F-31028 Toulouse, cedex 4, France 
\and 
CNRS, UMR 5187 F-31028, Toulouse, France
\and 
ASI Science Data Center, I-00044 Frascati (Rome), Italy
\and
Canadian Institute for Theoretical Astrophysics, University of
Toronto, 60 St. George Street, Toronto, ON M5S 3H8, Canada 
\and
Dipartimento di Fisica, Universita di Roma Tor Vergata, Rome, Italy
\and
Centre for Astrophysics Research, Science $\&$ Technology Institute,
University of Hertfordshire, Hatfield, AL10 9AB, UK
\and
INAF-IFSI - Via Fosso del Cavaliere 100, Rome, Italy}


\abstract
{Variations in the dust emissivity are critical for gas mass
  determinations derived from far-infrared observations, but also for
  separating dust foreground emission from the Cosmic Microwave
  Background (CMB). Hi-GAL observations allow us for the first time to study
  the dust emissivity variations in the inner regions of the Galactic
  plane at resolution below 1$\degr$.}
{We present maps of the emissivity spectral index derived from the combined \textit{Herschel} 
  PACS 160 $\mic$, SPIRE 250 $\mic$, 350 $\mic$, and 500 $\mic$ data, and the IRIS 100 $\mic$
  data, and we analyze the spatial variations of the spectral index as a
  function of dust temperature and wavelength in the two Science Demonstration Phase
  Hi-GAL fields, centered at l=30$\degr$ and
  l=59$\degr$.}
{ Applying two different methods, we determine both dust temperature and
 emissivity spectral index between 100 and 500 $\mic$, at an angular
 resolution ($\theta$) of 4$^{\prime}$.}
{Combining both fields, the results show variations of the emissivity spectral index in
  the range 1.8-2.6 for temperatures between 14 and 23 K. The median values of the spectral
  index are similar in both fields, i.e. 2.3 in the range
  100-500 $\mic$, while the median dust temperatures
 are equal to 19.1 K and 16.0 K in the l=30$\degr$ and
  l=59$\degr$ field, respectively. Statistically, we do not see any
  significant deviations in the 
  spectra from a power law emissivity between 100 and 500
  $\mic$. We confirm the existence of an inverse correlation between the
  emissivity spectral index and dust temperature, found in
  previous analyses.}
{}

\maketitle

\section{Introduction}
The large dust grains\citep[big grain component as defined in
][]{Desert90} dominate the total dust mass, as well as the
observed emission in the
far-infrared (FIR) domain \citep{Draine07}. They radiate in thermal equilibrium with
the interstellar radiation field, and their emission spectrum, assuming a fixed dust abundance and a single grain size,
is well approximated by
\begin{equation}
I_{\nu}(\lambda) \propto \epsilon_0 \left ( \frac{\lambda}{\lambda_0}
\right ) ^{-\beta} B_{\nu}(\lambda,T_d), 
\label{eq:Inu}
\end{equation}
where $\rm I_{\nu}(\lambda)$ is the specific intensity or brightness, $\rm B_{\nu}$ is the
Planck function, T$\rm _d$ is the dust temperature, $\rm \epsilon_0$
is the emissivity at wavelength $\rm \lambda_0$, and $\rm \beta$ is
the emissivity spectral index. 

Previous observations in the submillimeter (submm), at arcminute resolution, such as those by the
balloon-borne experiments PRONAOS
\citep[][]{Dupac03} and ARCHEOPS \citep[][]{Desert08} found evidence of an inverse
relationship between T$\rm _d$ and $\beta$ in various environments of the
interstellar medium (ISM). In PRONAOS data, variations
of the spectral index were observed in the range 2.4 to 0.8 for dust temperatures
between 11 and 80 K, whereas ARCHEOPS data showed a more pronounced
inverse relationship with $\beta$ values going from 4 to 1 between
7 and 27 K. Recently \citet{Veneziani10} highlighted a similar trend analyzing T$\rm _d$ and $\beta$ for eight high Galactic
latitude clouds, by combining IRAS, DIRBE, and WMAP data with BOOMERanG
observations. The $\beta$ values vary from 5 to 1 
in the temperature range 7-20 K, with a behavior similar to that
derived from ARCHEOPS data. These variations of $\beta$ with dust temperature
could be owing to intrinsic properties of amorphous dust grains, as
proposed by \citet{Meny07}, but the impact of the noise and the
temperature mixing along the line of sight (LOS) must be
carefully taken into consideration, before a physical interpretation
can be made.Variations of the apparent spectral index in a sample of cores in the Galactic plane have been observed
\citep{Hill06} using SCUBA and SIMBA.  However, because of the restricted
wavelength range of their observations, these authors could not attribute
them to temperature variations.
\begin{figure*}[!t]
\begin{center}
\includegraphics[width=14.7cm]{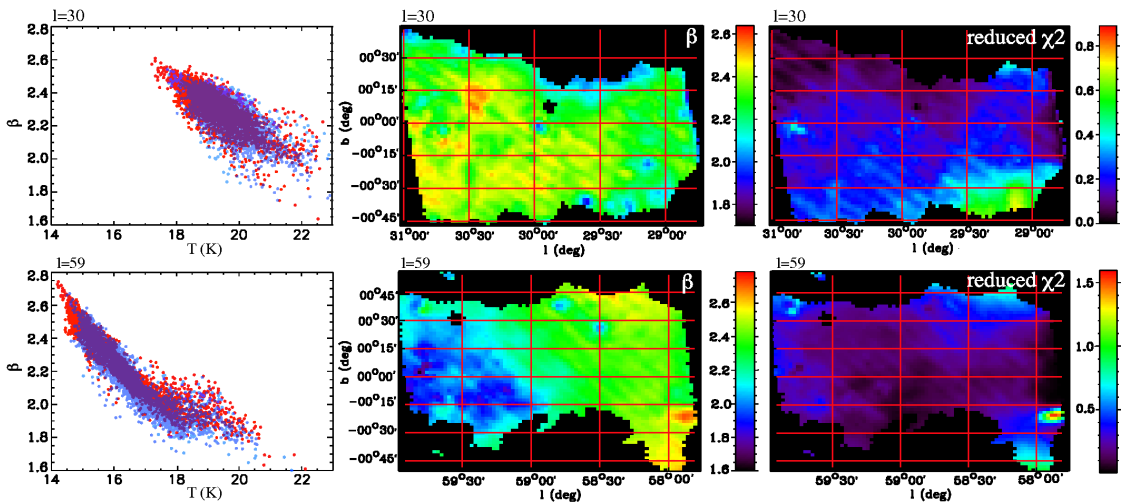}
\caption{Left: spectral index in the 100 to 500 $\mic$ range, versus dust
  temperature for each SDP field, combining IRIS 100 $\mic$ with PACS
  160 and SPIRE
  250, 350 and 500 $\mic$ data, using the least-square fit method in
  red and the MCMC method in blue. Middle and right: spatial distribution of
  the emissivity spectral index and reduced $\chi^2$, respectively, derived from the least-square fit
  method, for each SDP field.\label{fig_T_beta}}
\end{center}
\end{figure*}
Moreover, none of the previous analyses allowed the study of the
$\rm T_d-\beta$ correlation at arcminute scales in the inner regions of
the Galactic plane. 
Thanks to the \textit{Herschel} Hi-GAL data \citep[$\theta=37
^{\prime \prime}$, 2$\degr
\times$2$\degr$ maps, centered in the Galactic plane at l=30$\degr$
and l=59$\degr$, see][b]{Molinari10a}, we can now
extend for the first time this type of analysis to the inner Galactic plane, and on
  continuous fields of diffuse emission. An overview of the
\textit{Herschel} instruments is given in
\citet{Pilbratt10}.
In this paper we analyze emissivity variations between 100 and 500
$\mic$. We then derive both the dust temperature and spectral index
from Eq. (\ref{eq:Inu}), assuming a single temperature along the LOS. This assumption is relatively correct for the
l=59$\degr$ field, where there is less contamination along the LOS in
comparison with the l=30$\degr$ field. Indeed, the LOS toward the inner
l=30$\degr$ field crosses several regions with a presumably
wider range of temperatures.
An accurate temperature determination requires sampling both sides of the
emission peak.  The 70$\mic$ brightness is generally contaminated by
out-of-equilibrium emission from very small grains (VSGs) and cannot
be used to derive thermal dust temperature without an accurate
subtraction of this component. This contribution has been investigated
by \citet{Compiegne10}. Therefore we include the
  IRAS data at 100 $\mic$ in our analysis, for which we estimate an average VSG
    contamination of less than 10$\%$. This implies
degrading the original resolution of the Hi-GAL data by adopting
the IRAS angular resolution of 4$^{\prime}$. We note, as a
  consequence, an averaging effect within the resolution element,
  which also
  needs to be taken into account in the interpretation of the results
  of the present analysis.

One of the key aspects in the analysis of CMB
is the ability to separate its emission from the other astrophysical
foregrounds (including thermal dust) through multifrequency observations.
While it is not easy to achieve successful component separation over
the Galactic plane, determining the dust spectral index variations
across the sky can efficiently help in reducing the number of unknowns
in the problem \citep{ricciardi10}.
\section{Data}
We use the ROMAGAL \textit{Herschel} PACS and SPIRE maps described in
\citet{Traficante10}, combined with the IRIS \citep[Improved Reprocessing
of the IRAS Survey, see][]{MamD05} 100 $\mic$ data. Multiplicative factors (0.78,1.02, 1.05 and 0.94 at 160, 250,
350 and 500 $\mic$, respectively) have been applied to the data \citep{Poglitsch10,Griffin10,Swinyard10}. An absolute calibration accuracy of
20$\%$ and 15$\%$ for PACS and SPIRE has been adopted. We applied the
offsets given in \citet{Bernard10} \footnote{see their Table 1}. All maps have been
convolved with a Gaussian kernel, with a FWHM of 4$^{\prime}$,
i.e. equal to the IRIS 100 $\mic$
angular resolution. In addition, the maps have been rebinned on a uniform grid with
a pixel size of 1.65$^{\prime}$. The IRIS 100 $\mic$ calibration
uncertainty is taken to be 13.5$\%$ \citep[see][]{MamD05}.
\begin{figure*}[!t]
\begin{center}
\includegraphics[width=13.cm]{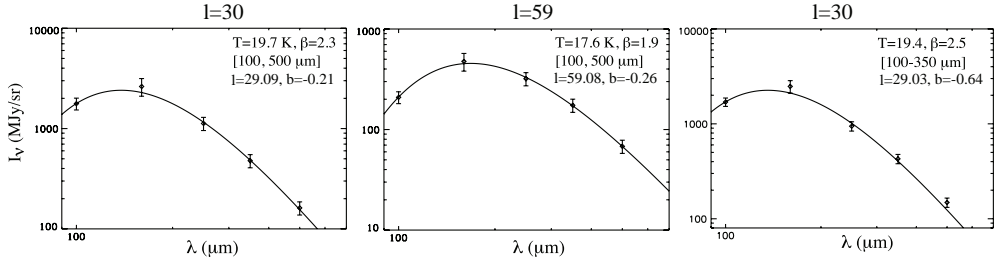}
\caption{Representative SEDs combining IRIS 100 $\mic$ with
  PACS 160 $\mic$ and SPIRE 250, 350 and 500 $\mic$, modeled by the gray-body
  law with T$\rm _d$ and $\beta$ derived from the least-square fit
  method between 100 and 500 $\mic$ (left and central panels), and
  between 100 and 350 $\mic$ (right panel).\label{fig_sed}}
\end{center}
\end{figure*}

\section{Variations of the emissivity spectral index with temperature}
\label{sec_em}
We use two different methods for the simultaneous derivation of the dust
temperature and the emissivity spectral index. Both allow us to
fit the data with a modified blackbody function (see Eq. \ref{eq:Inu}). In
the first method we perform an $\chi^2$ minimization, applying the same code as described in
\citet{Dupac01}, using the IDL least-square fit \texttt{curvefit}
function to deduce the T$\rm _d$ and $\beta$ parameters, as well as
their 1$\sigma$ uncertainties.
The second method estimates the best-fit parameters by looking for the
maximum likelihood, using a
Monte Carlo Markov Chain (MCMC) algorithm \citep{Lewis02} and
represents a bayesian approach to parameter
estimation. The posterior distribution for the parameters is sampled
by using the Metropolis-Hastings algorithm, and a maximum likelihood
estimate is derived jointly for $\beta$ and $T_d$. We chose a
  wide flat a priori probablity
  density of the parameters, i.e. 0 K$\rm <T_d<$60 K and -1$\rm
  <\beta<$5, in order not to constrain the fit results. A benefit of this
method is that it also recovers the joint posterior distribution of the
two estimated parameters. We make use of this feature to explore the
$\rm T_d$-$\beta$ correlation as explained below.

We fit the spectral energy distribution (SED) between 100 and 500
$\mic$ for each pixel of the maps, using both methods. The color
correction is computed iteratively in all channels. Results of the T$\rm _d$-$\beta$ determination for each
field are shown in Fig. \ref{fig_T_beta}. Only pixels with a
  surface brightness at
  500 $\mic$
  higher than 105 MJy/sr and 50 MJy/sr for the l=30$\degr$ and l=59$\degr$ fields,
respectively, have been considered. Both methods show a clear anti-correlation
T$\rm _d$-$\beta$ in each field. These two independent methods also highlight a good agreement. The $\beta$ values range 
from $\simeq2.6$ to $\simeq1.9$ for temperatures between 17.5-23 K for the l=30$\degr$ field, and from $\simeq2.7$ to $\simeq1.8$, between 14 and 21 K, for the l=59$\degr$ field. 
The assumption made of a single temperature along the LOS
may hold in the l=59$\degr$ field, but is certainly too simplistic
for the l=30$\degr$ field, since temperature variations along the LOS
are larger in the inner field \citep[see][]{Bernard10}. The analysis of the l=30$\degr$ field would certainly require an accurate
combination of temperatures and column densities. However, our methods do not allow 
the simultaneous fitting of a complex mixture of spectra. Even if the
l=30$\degr$ field is potentially characterized by several grain temperatures
along the LOS \citep[associated to Sagittarius, Scuttum-Crux and Perseus,
as described in][]{Bernard10}, and therefore could induce a spurious T$\rm _d$-$\beta$ inverse correlation
\citep{Masi95,Shetty09}, the l=59$\degr$ field, less contaminated by
various dust mixing effects, still highlights a pronounced behavior in
the T$\rm _d$-$\beta$ parameter space. Both fields, with a different mixture of temperatures, 
present a T$\rm _d$-$\beta$ anti-correlation. We therefore interpret
this as an indication that the mixture of temperatures is probably not
the dominant responsible effect for the observed anti-correlation. 
 
The spatial distribution of $\beta$ is shown in Fig. \ref{fig_T_beta}. Warmer regions, which most often correspond to bright
regions, show lower beta values. However,
the l=59$\degr$ field presents a gradient along the East-West direction,
which is not
observed in the PACS or SPIRE surface brightness maps. The stripes visible in the maps are
associated to low-level residual stripes in the IRIS
data. In the l=59$\degr$ field, $\beta$ values higher than 2.6 located at
l=57$\degr$48$^{\prime}$, b=-0$\degr$20$^{\prime}$ correspond
to one of the less active star-forming regions, although the PACS 160 data show an
artifact at this location, disclosed by a high $\chi^2$
(see Fig. \ref{fig_T_beta}). In the same field, the lowest $\beta$ values
($\leq 1.8$) are found for an HII region, with a bright source nearby at
l=59$\degr$38$^{\prime}$, b=+0$\degr$38$^{\prime}$. Representative
examples of SEDs with a gray-body fit are presented in Fig. \ref{fig_sed},
illustrating the agreement between the data and the model.

\begin{figure}[!t]
\begin{center}
\includegraphics[width=5.2cm]{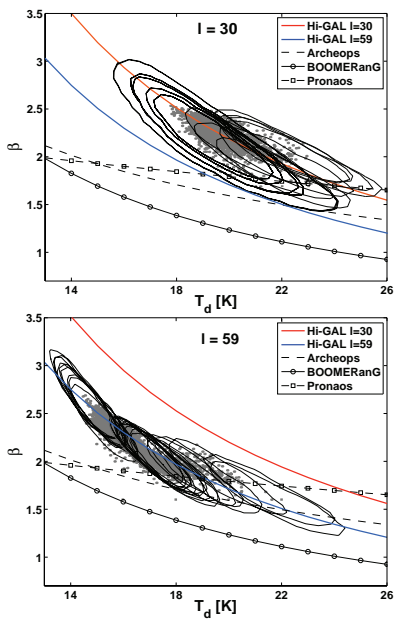}
\caption{T$\rm _d$-$\beta$ two-dimensional 68$\%$ contour posterior
  probabilities derived from the MCMC method, shown for $\simeq$15-25 pixels for the two SDP fields.The
  overplotted lines correspond to Hi-GAL (solid red and blue lines for
  the l=30$\degr$ and l=59$\degr$ fields, respectively), ARCHEOPS
  (dashed line), BOOMERanG (line with circles) and PRONAOS (line with
  squares) best-fits, respectively. The T$\rm _d$-$\beta$ relationship
  is estimated over all points within the contours. The gray points
  correspond to the T$\rm _d$-$\beta$ data points derived from the MCMC method (see Fig. 1,
  left panels).\label{fig_mcmc}}
\end{center}
\end{figure}

The two parameters T$\rm _d$-$\beta$ are degenerate in the parameter space,
creating a spurious inverse relation \citep{Shetty09} that has to be considered while
investigating if an intrinsic physical correlation indeed exists. In
order to properly estimate the relationship between T$\rm _d$ and
$\beta$, we have to take into account
the correlation that is shown by the spectral shape in
Eq. (\ref{eq:Inu}) and caused by calibration uncertainties. One way to include the effect of the degeneracy is 
to estimate this effect not only for the T$\rm _d$-$\beta$ best-fit values but for 
points included in the 68$\%$ contours of the two-dimensional posterior
probability obtained with the MCMC method. Within the 68$\%$  contours
the points are not uniformly distributed, but their density is higher
close to the maximum of the probability, increasing their weight, and lower close to the boundaries.
Assuming the model $\beta=A(T_d/20 \mathrm K)^{-\alpha}$, between 100 and 500$\mic$ in both fields, 
the fit is then performed on all the T$\rm _d$-$\beta$ couples inside
a posteriori probability contour of each pixel, chosen randomly, one for each pixel and weighted as described. 
We then take a point within the contour for each pixel of the map and
fit the trend to estimate the A and $\alpha$ parameters. We repeat
this procedure until we obtain a good sampling of the shape of the
distribution of posterior probabilities. This method has been
successfully tested on BOOMERanG data \citep{Veneziani10} and allows
us to include the effects of degeneracy in the error bars. The
resulting A and $\alpha$ ditributions are well approximated by
  Gaussian functions, even if a slight asymmetry is present. The center of the Gaussian is the more likely value, while the error is estimated by marginalizing over the other parameter. 
The best-fit values so obtained are\\
\begin{equation}
\beta=(2.19\pm0.25) \times \left (\frac{T_d}{20} \right)
^{-1.32\pm0.04}, \,l =30$\degr$
\end{equation}
\begin{equation}
\beta=(1.71\pm0.22) \times \left (\frac{T_d}{20} \right) ^{-1.33\pm0.04}, \, l =59$\degr$,
\end{equation}
Figure
\ref{fig_mcmc} shows the results of the fitting method over the
probability $68\%$ contours with only a few pixels out of $\sim3500$ contours for 
clarity. The best-fit Hi-GAL T$\rm _d$-$\beta$ relation is plotted, as
well as the ARCHEOPS, BOOMERanG, and PRONAOS ones for comparison. 
Our results highlight a steeper
T$\rm_d$-$\beta$ law in the inner Galactic plane in both fields than in external regions and high latitude clouds observed with ARCHEOPS and
BOOMEranG. Whereas the wavelength range covered
by PRONAOS is almost the same as that of \textit{Herschel}, the T$\rm_d$-$\beta$ relation 
($\beta=1/(\delta+\omega T_d)$) is flatter, with $\beta$ values
going from 2.0 to 1.7, in the
range 13-26 K. These
results could indicate different dust properties in the inner Galactic
plane, as compared to the outer
galaxy observed with ARCHEOPS, the solar neighborhood observed
with PRONAOS, and high latitude clouds observed with BOOMEranG. We
therefore think that these different T$\rm_d$-$\beta$ relations could
be caused by changes
in dust properties with the environments.

\section{Variations of the emissivity spectral index with wavelength}
In the previous section we have
found evidence of spectral index variations with temperature. As predicted by
\citet{Meny07}, a flattening of the emissivity
spectrum (and so a lower $\beta$) is expected for wavelengths longer
than 500 $\mic$ as a consequence of the internal structure of the
grains. Recently, \citet{Gordon10} highlighted
an emission excess at 500 $\mic$ in the Large Magellanic Cloud that could be of the same origin. However, this excess is of about
10$\%$, which is lower than the SPIRE calibration uncertainties of
15$\%$. The \textit{Herschel} data do not allow us to study emissivity variations at longer
wavelengths than 500 $\mic$. 

Although the SEDs in each field (left and
central panel of Fig. \ref{fig_sed}) do not provide evidence for any
departures from a power law, we analyze the possibility of an emissivity excess at 500
$\mic$. To properly perform this study, the temperature estimates need
to be done independently of the 500 $\mic$ data. Therefore we have
redone the T$\rm _d$ and $\beta$ determination with the least-square
fit method using only data between 100 and 350 $\mic$, and we compare
the emission at 500 $\mic$ with the value predicted by extrapolating
the emissivity power law between 100 and 350
$\mic$. The predicted values are
systematically lower than the data at 500 $\mic$ in both fields, by
16$\%$ in l=30$\degr$ and 13$\%$ in l=59$\degr$, 
which could favor the hypothesis of an emissivity excess at this
wavelength. These values are not significant compared to the
uncertainties on the SPIRE data. Moreover, we suspect that the PACS 160 $\mic$ data suffer from
calibration issues, as discussed in \citet{Bernard10}. The calibration
uncertainty at 160 $\mic$ does not impact the results of the
  global analysis described above, but here the limitation of the wavelength range (100-350 $\mic$)
removes some constraints on the fit, resulting in an overestimate with
respect to the data at 250 $\mic$, and an underestimate at 500 $\mic$
(see Fig. \ref{fig_sed}, right panel).  However, if we slightly shift down
the 160 $\mic$ data points, the apparent
underestimate at 500 $\mic$ would disappear. For this reason, we do not claim that this excess
is real and we think instead that the emissivity spectra are relatively
constant between 100 and 500 $\mic$ with similar median $\beta$ values in both fields,
i.e. 2.3, even if the median temperature is
statistically lower in the l=59$\degr$ field (T$\rm _d\simeq19.1$ K
for l=30$\degr$ and T$\rm _d\simeq16.0$ K for l=59$\degr$).
 
\section{Conclusions}
We investigated variations in the spectral index of the dust emissivity, with
temperature and wavelength, in the inner Galactic plane, using the new \textit{Herschel} observations in two Hi-GAL fields, centered at l=30$\degr$ and
  l=59$\degr$, acquired during the \textit{Herschel} Science
Demonstration phase, combined with the IRIS 100 $\mic$ data. We fitted the
SEDs for each pixel of the two fields with two independent methods, deriving
simultaneously the emissivity spectral index and the dust temperature
by adjusting a modified blackbody function to the data. The results are similar with both methods. Using a Monte Carlo Markov Chain algorithm 
method we computed the 68$\%$ likelihood contours for each point. We
find a T$\rm _d$-$\beta$ inverse correlation, with the local variation
going from 1.8 to 2.6 for temperatures between 14 and 23 K, shown
for the first time in the inner Galactic plane. Moreover, our results
indicate a different trend with respect to previous investigations based on BOOMERanG, ARCHEOPS, and
PRONOAS data, probably because of different dust
properties in the inner Galactic plane compared to other
environments. An extensive follow-up analysis will be required to take
into consideration the mixture of temperatures along the line of
sight. The median value of $\beta$ is similar in both fields, equal to 2.3, slightly
higher than the usual reference value of 2. We do not favor the hypothesis of an
emissivity excess at 500 $\mic$, as suggested for the Large
Magellanic Cloud. 
A complementary forthcoming study will combine
\textit{Planck} with \textit{Herschel} data to investigate possible changes in the dust emissivity
spectrum at wavelengths
larger than 500 $\mic$, as reported in previous studies of dust emission in our Galaxy. 

\section{Acknowledgments}
Data processing and maps production have been possible thanks to ASI generous support via contract I/038/080/0. M.V. is grateful to Davide Pietrobon for software support.

\end{document}